\documentclass[conference]{IEEEtran}
\IEEEoverridecommandlockouts
\usepackage{cite}
\usepackage{amsmath,amssymb,amsfonts}
\usepackage{algorithmic}
\usepackage{graphicx}
\usepackage{physics}
\usepackage{authblk}
\usepackage{booktabs,caption}
\usepackage[flushleft]{threeparttable}
\usepackage{hyperref}
\usepackage{textcomp}
\usepackage{xcolor}
\usepackage{siunitx}
\usepackage{multirow}
\usepackage{cuted}
\usepackage{subfiles}
\usepackage{amsmath}
\usepackage{amsthm}
\usepackage{capt-of}
\def\BibTeX{{\rm B\kern-.05em{\sc i\kern-.025em b}\kern-.08em
    T\kern-.1667em\lower.7ex\hbox{E}\kern-.125emX}}

\begin{document}

\title{Improving Quantum Classifier Performance in NISQ Computers by Voting Strategy from Ensemble Learning}


\author[1]{Ruiyang Qin}
\author[2]{Zhiding Liang}
\author[3]{Jinglei Cheng}
\author[4]{Peter Kogge}
\author[5]{Yiyu Shi}
\affil[1, 2, 4, 5]{University of Notre Dame}
\affil[3]{Purdue University}


\maketitle

\begin{abstract}

Due to the immense potential of quantum computers and the significant computing overhead required in machine learning applications, the variational quantum classifier (VQC) has received a lot of interest recently for image classification. The performance of VQC is jeopardized by the noise in Noisy Intermediate-Scale Quantum (NISQ) computers, which is a significant hurdle. It is crucial to remember that large error rates occur in quantum algorithms due to quantum decoherence and imprecision of quantum gates.
Previous studies have looked towards using ensemble learning in conventional computing to reduce quantum noise. We also point out that the simple average aggregation in classical ensemble learning may not work well for NISQ computers due to the unbalanced confidence distribution in VQC. Therefore, in this study, we suggest that ensemble quantum classifiers be optimized with plurality voting. On the MNIST dataset and IBM quantum computers, experiments are carried out. The results show that the suggested method can outperform state-of-the-art on two- and four-class classifications by up to 16.0\% and 6.1\% , respectively.
\end{abstract}


\section{Introduction}
In the past decade, machine learning models have achieved consistent success in many practical applications such as natural language processing \cite{young2018recent, sak2014long, qin2021ibert}, image classification \cite{he2016deep, luo2020open}, and medical diagnosis \cite{oktay2018attention}. However, as dataset size and model complexity increase, computation power in classical computing hardware has become the bottleneck. 
To alleviate the problem, various efforts have been made to leverage the power of quantum 
computers to speed up machine learning tasks, which have opend a research area known as 
quantum machine learning (QML). One of the most popular QML models is Variational 
Quantum Classifier (VQC) \cite{havlivcek2019supervised}, which deploys parameterized quantum circuits and trains them on the target classification tasks\cite{jiang2021co, chen2020variational}. 

Nevertheless, despite the great potentials of quantum computing, current Noisy Intermediate-Scale Quantum (NISQ) computers are extremely sensitive to their surrounding environments and risk losing their quantum state due to quantum decoherence, which contributes to high quantum noise.
And it is theoretically impossible to accurately predict or eliminate quantum noise\cite{zlokapa2020deep} in NISQ computers. 
As a result, numerous studies have been done in the literature to reduce the noise by constructing noise-resistant circuits\cite{wang2022quantumnas, wang2022quantumnat, liang2021can} or controlling lower-level quantum hardware \cite{liang2022pan, liang2022variational, cheng2020accqoc}. 
Many of these methods are task-specific, i.e., they are developed with the implementation of certain quantum algorithms in mind.

\begin{figure}
    \centering
    \includegraphics[scale = 0.21]{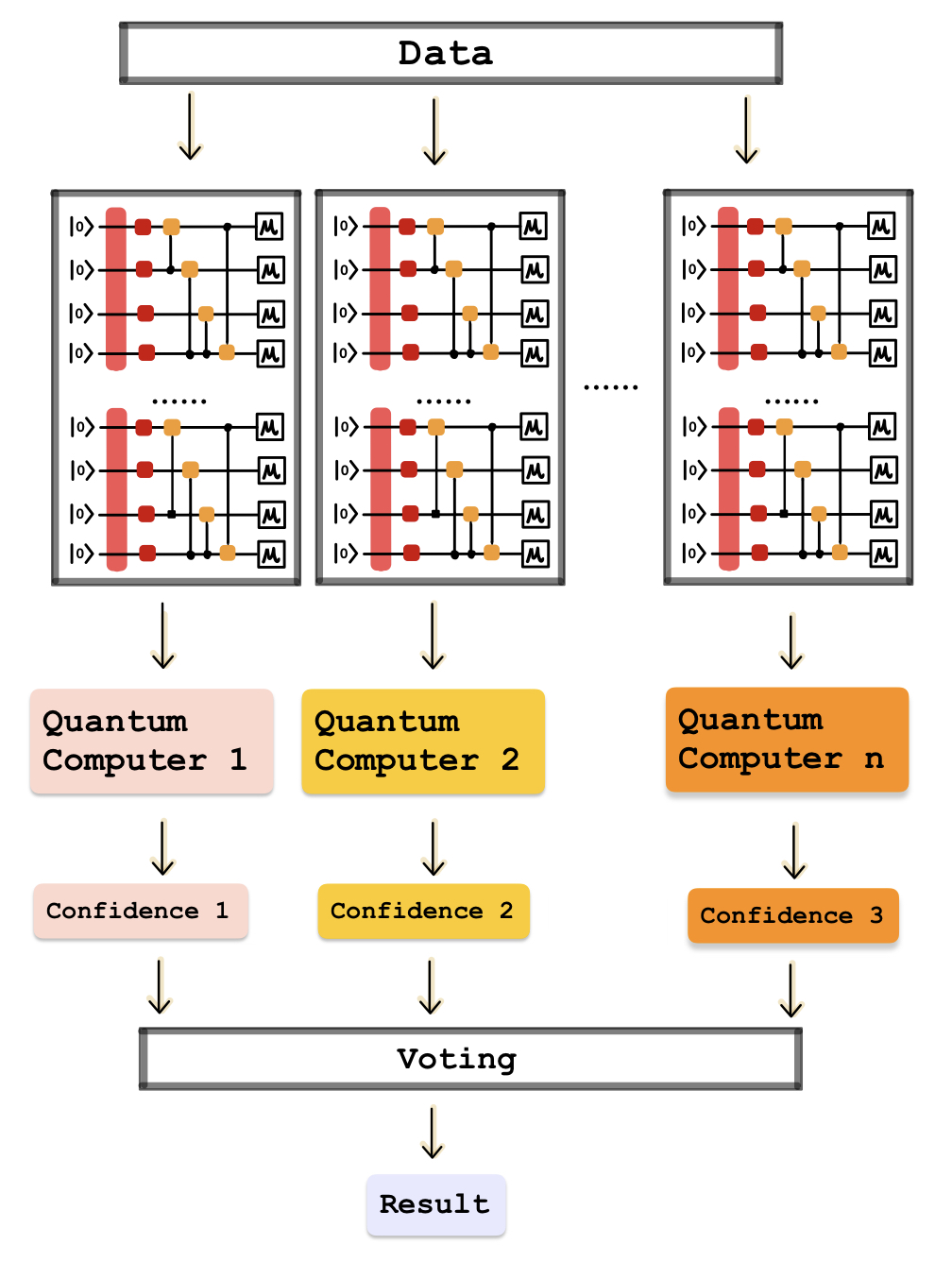}
    \caption{Conceptual illustration of EQV. Different variational quantum circuits and quantum computers are used to produce the confidence numbers for input data. A plurality voting system is proposed to generate the final predictions}
    \label{fig1}
\end{figure}

More specifically, for VQC, current noise mitigation techniques include mitigating noise-induced gradient vanishment\cite{wu2021mitigating}, adopting hybrid optimization\cite{gentini2020noise}, 
pre-processing data\cite{larose2020robust}, and developing kernel matrix\cite{pillay2021implementing}. 
Recently, QUILT\cite{silver2022quilt} was proposed for VQC tasks. It was inspired by the idea of ensemble learning, where a mixture of various models make a prediction collectively for greater accuracy. 
To generate the final prediction, it deploys several VQCs and averages their output confidence, which is defined as the possibilities that the prediction being correct. Although quantum noise is unpredictable, as will be shown in Section II, we note that the impact of wrong predictions is typically much more than that of correct predictions. As a result, utilizing averaged confidence may result in incorrect prediction.

In this paper, we present EQV, an \textbf{E}nsemble \textbf{Q}uantum classifiers with plurality \textbf{V}oting, to address this issue. 
As illustrated in Fig.~\ref{fig1}, EQV deploys numerous VQCs onto multiple quantum computers and integrates the outcoming results through plurality voting to produce the final prediction. If the number of quantum computers producing accurate predictions predominate, such an approach can successfully eliminate inaccurate predictions, even if they have very low confidence numbers.
Experimental results on real-world quantum computers with MNIST dataset demonstrate that EQV can increase the accuracy by 16.0\% and 6.1\% over the state-of-the-art method for two-class and four-class classification tasks respectively.

The rest of the paper is structured as follows. Section II provides background information and introduces the motivation for our work. The detailed framework of quantum ensemble learning for VQC is articulated in Section III. Section IV contains the experimental results as well as the concluding remarks.

\section{Background and Motivation}
\subsection{Background}

The quantum bit (qubit) is the fundamental building block of quantum information. Similar to how a conventional bit may store either a 0 or 1, a qubit can also be used to store information. The states of a qubit can be represented by two vectors: $\ket{0}$ and $\ket{1}$. The linear combination of these two state vectors is superposition, which can be represented as
\begin{equation}
    \ket{\psi} = \alpha\ket{0} + \beta\ket{1},
\end{equation}
where $\alpha$ and $\beta$ should satisfy ${|\alpha|^2} + {|\beta|}^2 = 1$. 

Besides superposition, qubits can also be entangled, which is impossible for classical bits.
Using two-qubit gates such as CNOT gates to connect two qubits is a typical method for entangling qubits. 

A quantum circuit is a collection of various quantum gates that can perform quantum operations efficiently. Parameterized quantum circuit is a type of quantum circuit that can be parameterized to enable trainability by changing the angles of the rotation gates. We are able to obtain results that are analogous to those obtained from training a classical neural network if we design and train the variational quantum classifier. For instance, variational quantum classifiers (VQC) are adequate for solving image classification problems in an effective manner~\cite{li2021vsql, maheshwari2021variational}.

The leading quantum computers in the NISQ era contain around one hundred qubits, but they are not advanced enough to achieve fault-tolerance nor large enough to demonstrate quantum supremacy on practical problems. The performance of these computers is restricted by their high quantum noise. Therefore, powerful QML applications such as quantum neural network \cite{jiang2021co, liang2022hybrid}, which is believed to better simulate human neurons than classical neural networks, cannot be implemented.

\subsection{Motivation and Related Work}

The idea of quantum ensemble learning was brought up in 2017, where the ensemble corresponds to state preparation routines, and the quantum classifiers can be evaluated in parallel \cite{schuld2018quantum}. 
Schuld et al theoretically proves the feasibility of quantum ensemble learning. In another work, Macaluso et al \cite{macaluso2020quantum} propose a quantum algorithm which takes the advantages of quantum superposition, entanglement and interference to build an ensemble classification. Macaluso et al \cite{macaluso2020quantum} achieves quantum ensemble learning by using the bagging strategy \cite{ganaie2021ensemble}, and it mainly discusses how to apply classical ensemble learning methods to quantum computing. Chen et al \cite{chen2016quantum} combines quantum ensemble learning with supervised learning by recasting quantum ensemble classification as a supervised quantum learning problem, and using a sampling-based learning control to present quantum discrimination. 
There are also researchers working on building an exponentially larger ensemble size of classifiers, aiming to explore the scalability problem \cite{araujo2020quantum}. In hierarchical quantum classifiers \cite{grant2018hierarchical}, researchers use the combination of different quantum binary classifiers to complete classification tasks. 

Most recently, QUILT \cite{silver2022quilt} deploys five core classifiers and $N$ binary classifiers for N-class classification tasks. In QUILT, accuracy-based weights are used to aggregate outputs from core classifiers. 
The binary classifiers in QUILT are One-Vs-All classifiers, which are trained to discriminate one class from other classes. In QUILT, binary classifiers are used to tell apart two outputs with lowest confidence numbers. For example, if the lowest confidence classes are $one$ and $five$, the binary classifier will be adopted to make the final decision.

\begin{figure}
    \centering
    \includegraphics[scale = 0.49]{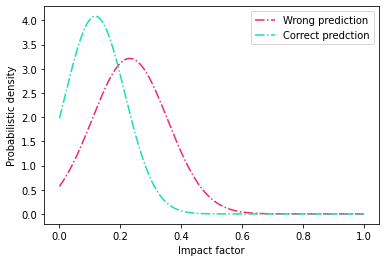}
    \caption{Probability density vs. Impact factor for two-class classification with two-qubit ansatz. It illustrates that wrong prediction has higher impact than correct prediction, if a simple averaging aggregation is adopted.}

    \label{fig_two_class}
\end{figure}

We conduct a toy experiment and present its results in Fig.~\ref{fig_two_class} to demonstrate the distribution of ``impact numbers'' for wrong and correct prediction. 
For an input image, two numbers are generated by multiple models to determine the class of the input image. 
By comparing these two numbers, the larger numbers indicate the classes predicted by these models. 
To evaluate how the numbers will take effect in the average, we define an ``impact factor'' for predictions. For predictions, the factor is defined as the difference between two confidence numbers generated by the models. 
When the confidence number are averaged in ensemble learning models, the factor number will determine the ``weight'' of this model. The larger the factor number is, the more impact this single model will have on the ensemble learning model. And in the experiment, we find out that if the predictions are wrong, as shown in Fig.~\ref{fig_two_class}, their impact factors are generally larger than the impact factors of correct predictions. 
For example, we have three binary classification models and each model will take an image of digit of 1 as its input. 
The model 1 generates confidence numbers of class 1 and class 0 as \{0.1, 0.9\},
the model 2 generates confidence numbers of class 1 and class 0 as \{0.6, 0.4\},
and the model 3 generates confidence numbers of class 1 and class 0 as \{0.55, 0.45\}. So the impact factors are 0.8 for model 1, 0.2 for model 2, and 0.1 for model 3. Model 2 and model 3 should generate the correct classification results. However, through simple averaging method, the ensembled model has confidence of 1 as 0.417 and confidence of 0 as 0.58. Therefore, the ensembled model will choose digit 0 as its final prediction class. Due to the large impact factor, model 1 governs the ensemble model's prediction. On the contrary, since we know the predictions of the three models as 0, 1, 1, we can make the final prediction as digit 1 because more models ``choose'' digit 1 as the results.

As shown in Table~\ref{table:1}, different quantum computers have different error rates. The impact of quantum noise on quantum computers are also difficult to predict \cite{resch2021benchmarking}. Running all quantum classifiers on the same quantum computer may lead to low performance if this quantum computer receives high quantum noise. To eliminate this bias introduced by quantum noise, we propose to run experiments on different quantum computers.
We propose a quantum ensemble model in this paper that trains quantum classifiers across multiple quantum computers. Instead of simply averaging the outputs of these classifiers, we generate the final output using a voting strategy called plurality voting. As a result, our work avoids over-reliance on a single quantum computer and mitigates the biased impact of low-accuracy quantum classifiers.

\section{Architecture: EQV}

\subsection{Overview}

As shown in Fig.~\ref{fig1}, EQV uses many variational quantum classifiers (VQCs) to train on the same task. EQV makes copies of VQCs and deploys them onto several quantum computers.
Each VQC will generate its own predictions.  And the confidence of all VQCs' outputs are different. In this work, we use intermediate confidence to represent each VQC's output.
As shown in Fig.~\ref{fig1}, our voting strategy collects all intermediate confidence and generates the final output.
Instead of averaging all intermediate confidence and adding it up for voting, it uses a plurality voting approach to combine all inputs (intermediate confidence). 

We show the behaviors of EQV using the two-class classification as an example. In this illustration, we use EQV to categorize the digits $one$ and $zero$.
Assuming there are five two-qubit quantum circuit–based classifiers for quantum systems. The five quantum classifiers will provide their predictions given an image. Three quantum classifiers indicate the given image being a digit of $one$ if they individually generate the confidence numbers as 0.57, 0.63, 0.38, 0.27, and 0.61. The digit $one$ will be chosen as the final result based on plurality voting. But for QUILT aggregation technique, we will have the averaged confidence number of 0.492, resulting in the wrong prediction. In this case, voting is capable of making a reliable prediction without being affected by the two low-confidence outcomes (0.38 and 0.27).

\subsection{Ensemble the Outputs: Plurality Voting}

Voting is a way to aggregate outputs from various VQCs. In a classification task, a classifier predicts a class label from a set of class labels. For example, a classifier predicts the input image has 0.58 as confidence number to be $label_1$, then the predicted label of this image is $label_1$. In voting, this prediction represents one classifier's vote. In the final output, the class label that obtains the most votes represents the final prediction.
This voting strategy is called plurality voting. The final prediction does not need to receive more than 50\% votes. 
For instance, there are three classifiers for classifying data into two classes (class 1 and class 2). Upon receiving an input, classifiers
output confidence on class 1 (correct class): \{0.6, 0.55, 0.1\} and 
output confidence on class 2:                 \{0.4, 0.45, 0.9\}. Class 1 receives the most votes in the three classes. 
Thus, EQV makes accurate predictions.
Again, the average level of confidence for class 1 is 0.4167, while for class 2 it is 0.5833. Using average as the aggregate method, the final prediction will be class 2, which is incorrect.

In contrast to the suggested plurality voting, existing quantum ensemble learning methods, as explained in Section II, use average aggregation, which adds all outputs and computes the average of the sum. This is similar to how classical machine learning works, where an ensemble output often averages the outcomes of numerous models trained to accomplish the same task.
Because some properties may be captured better by one model than by another, the final result should be superior to any single output. However, quantum noise causes certain outputs to perform far worse than others. In QML, average output aggregation may reduce confidence. 

We present an example below to demonstrate plurality voting. There are four labels to choose from. The correct label for a given image is $c_1$. In this case, there are six incorrect predictions. When employing average aggregation, the final prediction may be incorrect since incorrect predictions have a bigger impact on the average number. Our voting strategy, on the other hand, can yield the correct prediction.
\begin{equation}
\begin{aligned}
    label = \{c_1, c_3, c_6, c_9\} \\
    dataset = \{img_i, i = 1\sim300\} \\ 
    img_i: \{c_1 * 3, c_3 * 2, c_9 * 2, c_6 * 2\} \\ 
    Voting(img_i) \Rightarrow c_1
\end{aligned}
\end{equation}
This example demonstrates that even if more than half of quantum classifiers make incorrect predictions due to quantum noise, voting can still produce the correct prediction.

\section{Experiments}

\subsection{Experimental Setup}
\begin{figure}
    \centering
    \includegraphics[scale = 0.12]{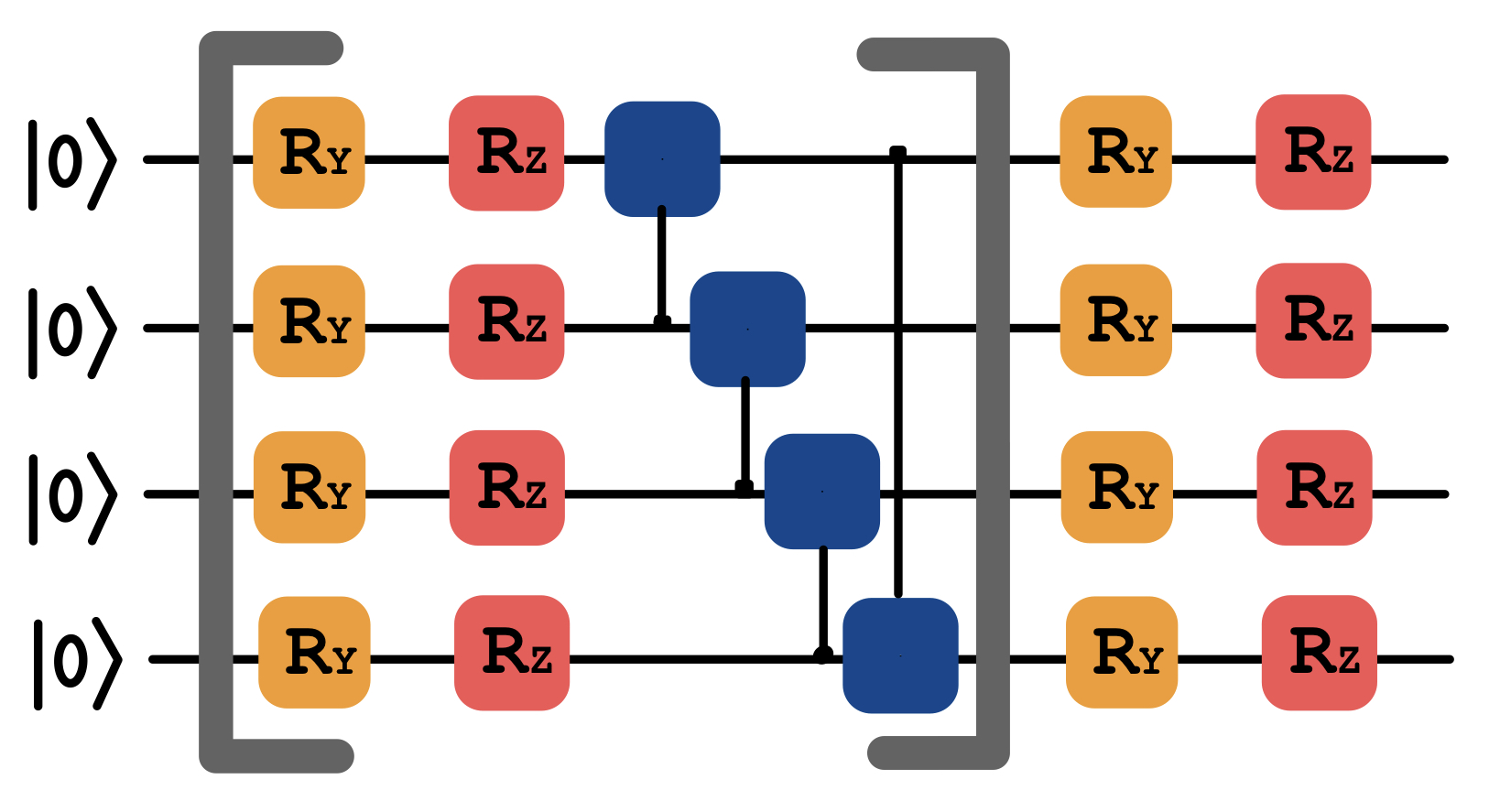}
    \caption{The ansatz we use as the basic ansatz is from the idea of Hardware-Efficient Ansatz \cite{kandala2017hardware}, where one-qubit gates are applied to all qubits and two-qubit gates are applied to all possible connections between physical qubits.}
    \label{fig3}
\end{figure}

We undertake two experiments to evaluate EQV's performance: a two-class classification experiment and a four-class classification experiment.
The training set for two-class classification consists of 300 MNIST-2 images, and the test set consists of 30 MNIST-2 images. The two classes are made up of digits $one$ and $nine$. The training set for four-class classification consists of 300 MNIST-4 images, and the test set consists of 30 MNIST-4 images. The four categories are as follows: digits $one, four, seven\ and\ nine$. A batch size of 100 is used. Three training subsets are established in this arrangement. 
We construct all quantum classifiers based on the quantum circuit shown as Fig.~\ref{fig3}. 
We position rotating gates at the beginning of the quantum circuit. CNOT gates are put after the leftmost rotation gates, and more rotation gates may follow CNOT gates. If their CNOT gates or rotation gates are arranged differently, two quantum circuits with the same number of qubits can be considered variations of one another.
However, a quantum circuit with three qubits cannot be a variation of a circuit with two qubits. Likewise, a quantum circuit lacking CNOT gates or rotation gates cannot be called a version of a quantum circuit containing both type gates.
In Fig.~\ref{fig3}, The block enclosed by the grey edges is known as an entangled unitary. No matter how many qubits a quantum circuit has, an entangled unitary should have rotation gates and CNOT gates entangle all qubits. To change an entangled unitary, we can position CNOT gates in different positions, thus entangling distinct qubits. (entangling qubit1 $\&$ qubit2 $\rightarrow$ entangling qubit1 $\&$ qubit3). For dataset, we split it into many batches. For example, if the batch size is 128, then 128 images will be convert into tensors which will be considered as an input. By reshaping, the input will be embedded into rotation gates

As shown in Fig.~\ref{fig1}, Different variants are bundled together and deployed on a single quantum computer. However, the same variants will be deployed to different quantum computers. If three variants are constructed, three VQCs will be performed successively in a quantum computer. These variants' copies will be transferred to a different quantum computer. For example, if we have four variants, each with two copies, we will send four sets of VQCs to three separate quantum computers.

We run the experiments on five IBM quantum computers, $ibmq\_lima, ibmq\_quito, ibmq\_belem, ibmq\_nairobi$ and $ imbq\_oslo$, with their properties shown in Table~\ref{table:1}. 
Since our quantum circuits consist of four qubits, we do the majority of our work on quantum computers with five qubits. We also present tests on two quantum computers with seven qubits.
\begin{table}[h!]
    \centering
    \begin{tabular}{p{1.5cm}p{0.7cm}p{1.2cm}p{1.1cm}p{0.4cm}p{0.7cm}}\toprule
         Quantum        &  qubit    & readout error & CNOT error    & QV & shots\\ \midrule
         ibmq\_lima     &  5        & $2.734^{-2}$  & $1.166^{-2}$  & 8  & 1024 \\
         ibmq\_quito    &  5        & $4.714^{-2}$  & $9.675^{-3}$  & 16 & 1024 \\
         ibmq\_belem    &  5        & $3.080^{-2}$  & $6.176^{-2}$  & 16 & 1024 \\
         ibm\_nairobi  &  7        & $4.599^{-2}$  & $1.015^{-2}$  & 32 & 1024 \\
         ibm\_oslo      &  7        & $2.411^{-2}$  & $1.111^{-2}$  & 32 & 1024 \\ \bottomrule
    \end{tabular}
    \caption{Five IBM quantum computers are utilized for this project. QV stands for quantum volume, which is a single-number metric that may be measured using a specific protocol. A single execution of a quantum algorithm on a quantum computer is referred to as a shot. Thus, the number of repetitions for each quantum algorithm is denoted by shots. }
    \label{table:1}
\end{table}

Voting is used during the testing process. The test set is distributed to each VQC, and the outputs of all VQCs are gathered and forwarded to the voting system. Using a plurality voting approach, the final outcome is determined. Together, the final output and test set labels can be used to calculate the accuracy.

\subsection{Experimental Results}

The quantum circuits used in the two experiments mentioned in the previous section should have the same number of qubits. Another experiment is conducted to compare the performance of quantum classifiers with varying numbers of qubits.
As shown in Table~\ref{tabII}, We use VQCs with two, four, and six qubits to conduct two-class classifications. A 2-qubit VQC, for example, has two entangled qubits, one CNOT gate followed by two rotation gates for each qubit. It can be viewed as a ``half'' of the ansatz depicted in Fig.~\ref{fig3}.

\begin{table}[h!]
    \centering
    \begin{tabular}{p{1.3cm}p{1.6cm}p{1.6cm}p{1.6cm}}\toprule
        Machine      & 2-qubit          & 4-qubit          & 6-qubit         \\ \midrule
        ibmq\_lima   & 0.65 $\pm$ 0.032 & 0.79 $\pm$ 0.025 & -\\
        ibmq\_quito  & 0.67 $\pm$ 0.019 & 0.82 $\pm$ 0.021 & -\\
        ibmq\_belem  & 0.64 $\pm$ 0.030 & 0.81 $\pm$ 0.045 & -\\
        ibm\_oslo    & 0.62 $\pm$ 0.038 & 0.77 $\pm$ 0.028 & 0.83 $\pm$ 0.017\\
        ibm\_nairobi & 0.61 $\pm$ 0.034 & 0.80 $\pm$ 0.037 & 0.81 $\pm$ 0.025\\ \bottomrule
    \end{tabular}
    \caption{Two-class classification accuracy on different quantum computers based on different number of qubits. For 2-qubit setting, a VQC with two qubits is involved. The VQC is designed based on the description of experimental setup. For 4-qubit and 6-qubit settings, four-qubit and six-qubit VQC based on the same designs. The quantum computers ibmq\_lima, ibmq\_quito, and ibmq\_belem have only five qubits and they cannot be used by six-qubit VQCs.]
    }
    \label{tabII}
\end{table}

We use five quantum computers shown in Table~\ref{tabII}, the first three are five-qubit quantum computers, and the last two are seven-qubit. In two-qubit setting, the three five-qubit quantum computers are slightly better than the rest two. They are have around 65\% of accuracy. In four-qubit setting, the five quantum computers in general have around 80\% of accuracy, which is much higher than that in two-qubit setting. Only \textit{ibm\_oslo} and \textit{ibm\_nairobi} can perform six-qubit setting of experiment, their performance in average is slightly better than that in four-qubit setting. 

What we find from this experiment:
\begin{itemize}
    \item For classification tasks, the VQC with more qubits might achieve higher performance. 
    \item Four qubits can be used to handle two-class classification with a comparatively high accuracy.
\end{itemize}

EQV is evaluated on two-class and four-class classification. In both experiments, we use four-qubit VQCs. For each experiment, we set five different ensemble size for EQV. 
Ensemble size in our work can be seen as the number of VQCs. We evaluate EQV in ensemble size of 3, 5, 7, 9, and 11. For ensemble size of 3, there are three variants and they all have no copies. For ensemble size of 5, three variants are created, and randomly pick two of three to make one copy of each. For ensemble of 7, three variants are created as well, two of them will have one copy, and one of them will have two copies. Similar rules apply to ensemble size of 9 and 11.

\begin{figure}
    \centering
    \includegraphics[scale = 0.4]{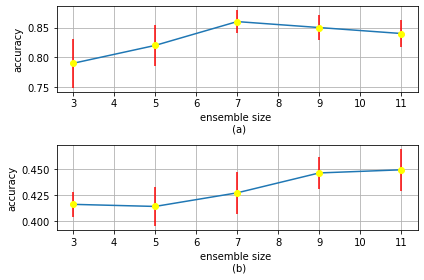}
    \caption{Accuracy vs. ensemble size for (a) MNIST-2 and (b) MNIST-4 classification using four-qubit ansatz. We can tell that for the simpler MNIST-2 task, it is earlier to reach accuracy satuation as the ensemble size increases.}
    \label{fig5}
\end{figure}

We perform EQV based on three IBM quantum computers: ibmq\_lima, ibmq\_belem, and ibmq\_quito. As shown in Fig.~\ref{fig5}, in two-class classification, ensemble size of 7 has the highest accuracy (87\%). In four-class classification, ensemble size of 11 has the highest accuracy (45\%). We use ensemble size of 7 for two-class classification and ensemble size of 11 for four-class classification to run single VQC experiment (One VQC on one quantum computer without ensemble and voting).

\begin{figure}
    \centering
    \includegraphics[scale = 0.5]{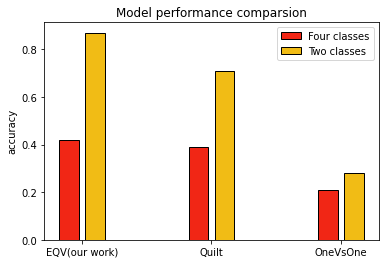}
    \caption{Accuracy comparison between EQV and state-of-the-art quantum ensemble learning schemes QUILT and One-Vs-One using 4-qubit ansatz on MNIST-4.}
    \label{fig6}
\end{figure}

The performance of EQV shown in Fig.~\ref{fig6}, where two-class classification has accuracy of 87\% and four-class classification has accuracy of 45\%. As Table III shown, while the performance of our work in real quantum computer is lower than simulation performance (0.91 for two-class classification and 0.71 for four-class classification), the performance of our work is still higher than QUILT \cite{silver2022quilt} and One-Vs-One \cite{romera2013one}.

As shown in Table III, We compare the performance of EQV to that of the same ansatz on a single quantum computer. If we just run the model on the same quantum computers, the performance in two-class and four-class classification is no better than EQV, which runs the model on three quantum computers. However, due to quantum noise, the performance of EQV is worse than that of simulation. Furthermore, the accuracy drop from simulation to EQV is greater in four-class classification than in two-class classification. It demonstrates that the more qubits involved, the greater the probability of being influenced by quantum noise.

\begin{table}[h!]
    \centering
    \begin{tabular}{p{1.9cm}p{2cm}p{2cm}} \toprule
        Quantum         & two-class     & four-class     \\ \midrule
        EQV(our work)   & 0.87 $\pm$ 0.020 & 0.451 $\pm$ 0.027 \\ \midrule
        ibmq\_lima      & 0.81 $\pm$ 0.038 & 0.416 $\pm$ 0.046 \\
        ibmq\_quito     & 0.83 $\pm$ 0.052 & 0.413 $\pm$ 0.057 \\
        ibmq\_belem     & 0.78 $\pm$ 0.027 & 0.406 $\pm$ 0.021 \\ \midrule
        simulation      & 0.91             & 0.71              \\ \bottomrule
    \end{tabular}
    \caption{EQV performance over vqc on single quantum computer performance. EQV represents our work's performance using ibmq\_lima, ibmq\_quito, and ibmq\_belem together to train the model. These three quantum computers are also used to train vqc individually. Using same vqc (ansatz), we also evaluation its simulation performance. We use MNIST-2 as our dataset}
    \label{tabIII}
\end{table}

In two-class classification tasks shown as Fig.~\ref{fig6}, the performance of our work (87\%) is better than QUILT (71\%) and better than One-Vs-One (28\%). In four-class classification tasks shown as Fig.~\ref{fig6}, the performance of our work (45.1\%) is better than QUILT (39\%) and better than One-Vs-One (21\%). 

\section{Conclusion and Discussion}

We propose EQV, a quantum ensemble learning framework with the capability of noise mitigation. We use plurality voting strategy from classical ensemble learning to integrate the outputs from all sub-quantum classifiers. We study the impact of information loss based on this voting strategy. By deploying quantum classifiers into different quantum computers, combining with this voting strategy, our work achieves better performance in accuracy and stability than the state of art works under the same settings. On real NISQ machines, we are able to achieve up to 16.0\% and 6.1\% higher accuracy compared with state-of-the-art on two- and four-class classifications.

In quantum ensemble learning, we observe the unbalanced confidence distribution of correct and incorrect predictions provided by quantum classifiers from the distributions shown in Fig.~\ref{fig_two_class}. We go over instances when an unbalanced distribution prevents average aggregation from generating correct results. In addition to quantum noise, the unbalanced distribution may result from structure of quantum circuit and its data processing. In our work, we find that plurality voting can be utilized to address the unbalanced distribution. 

Multiple quantum classifiers are utilized to perform the same task in this study. Each quantum classifier can also be used to perform a subset of the overall task and incorporate the results into the output. Due to the differences in their underlying hardware and structural makeup, quantum classifiers differ from traditional machine learning classifiers. In addition to predictions, a quantum classifier's output may also include information regarding outputs other than prediction and accuracy, such as the effects of quantum noise. Although such information cannot be quantified, it will influence the result in other ways, such as the confidence distribution. In this manner, the outputs of quantum classifiers must be independently analyzed. And we propose to do so with EQV.

\bibliographystyle{IEEEtran}
\bibliography{main}

\end{document}